\documentclass[aip,rsi,reprint,graphicx]{revtex4-1} 
\usepackage{amsmath} 
\usepackage{amsthm} 
\usepackage{amssymb}	
\usepackage{graphicx}
\usepackage{dcolumn}
\usepackage{bm}
\usepackage{hyperref}
\usepackage{sidecap}
\usepackage{float}
\usepackage{color}

\draft

\begin{document}


\title{Construction of a $^3$He magnetic force microscope with a vector magnet}

\author{Jinho Yang}
\affiliation{Center for Artificial Low Dimensional Electronic Systems, Institute for Basic Science, 77 Cheongam-Ro, Nam-Gu, Pohang 790-784, Korea}
\affiliation{Department of Physics, Pohang University of Science and Technology, Pohang 790-784, Korea}

\author{Ilkyu Yang}
\affiliation{Center for Artificial Low Dimensional Electronic Systems, Institute for Basic Science, 77 Cheongam-Ro, Nam-Gu, Pohang 790-784, Korea}

\author{Yun Won Kim}
 \altaffiliation[Present Address: ]{Division of Physical Metrology, Korea Research Institute of Standards and Science, Daejeon 305-340, Korea}
\affiliation{Center for Artificial Low Dimensional Electronic Systems, Institute for Basic Science, 77 Cheongam-Ro, Nam-Gu, Pohang 790-784, Korea}

\author{Dongwoo Shin}
\affiliation{Center for Artificial Low Dimensional Electronic Systems, Institute for Basic Science, 77 Cheongam-Ro, Nam-Gu, Pohang 790-784, Korea}
\affiliation{Department of Physics, Pohang University of Science and Technology, Pohang 790-784, Korea}

\author{Juyoung Jeong}
\affiliation{Center for Artificial Low Dimensional Electronic Systems, Institute for Basic Science, 77 Cheongam-Ro, Nam-Gu, Pohang 790-784, Korea}
\affiliation{Department of Physics, Pohang University of Science and Technology, Pohang 790-784, Korea}

\author{Dirk Wulferding}
\affiliation{Center for Artificial Low Dimensional Electronic Systems, Institute for Basic Science, 77 Cheongam-Ro, Nam-Gu, Pohang 790-784, Korea}

\author{Han Woong Yeom}
\affiliation{Center for Artificial Low Dimensional Electronic Systems, Institute for Basic Science, 77 Cheongam-Ro, Nam-Gu, Pohang 790-784, Korea}
\affiliation{Department of Physics, Pohang University of Science and Technology, Pohang 790-784, Korea}

\author{Jeehoon Kim}
 \email{jeehoon@postech.ac.kr}
\affiliation{Center for Artificial Low Dimensional Electronic Systems, Institute for Basic Science, 77 Cheongam-Ro, Nam-Gu, Pohang 790-784, Korea}
\affiliation{Department of Physics, Pohang University of Science and Technology, Pohang 790-784, Korea}

\date{\today}

\begin{abstract}
We constructed a $^3$He magnetic force microscope operating at the base temperature of 300 mK under a vector magnetic field of 2-2-9 T in the $x-y-z$ direction. Fiber optic interferometry as a detection scheme is employed in which two home-built fiber walkers are used for the alignment between the cantilever and the optical fiber. The noise level of the laser interferometer is close to its thermodynamic limit. The capabilities of the sub-Kelvin and vector field are demonstrated by imaging the coexistence of magnetism and superconductivity in a ferromagnetic superconductor (ErNi$_2$B$_2$C) at $T$=500 mK and by probing a dipole shape of a single Abrikosov vortex with an in-plane tip magnetization. 
\end{abstract}


\maketitle

\section{Introduction}
Atomic force microscopy (AFM) is one of the most versatile imaging techniques to investigate surface science at the nanoscale. Magnetic force microscopy (MFM),~\cite{Martin} a variant of AFM with a magnetized tip, plays an important role in studying the formation of magnetic domains  in magnetic materials,~\cite{Schwarz2, Liebmann, LSMO} vortex dynamics in superconductors,~\cite{Moser, Volodin, Schwarz, Roseman2, Pi, Luan, EWJ, Nb} and exotic spin structures such as skyrmions.~\cite{Milde} So far, several home-built low-temperature MFMs have been operated down to the $^4$He temperature within a single axis magnet.~\cite{Hug, Euler, Roseman, Chuang, Liebmann, Nazaretski} On the other hand, MFM apparatus operating under high demanding conditions, such as sub-Kelvin temperatures and vector magnetic fields, are rare,~\cite{Pelekhov, Ozgur} although they are useful for understanding unconventional (e.g., heavy fermion) superconductivity~\cite{Thalmeier} and quantum/molecular magnetism at the nanoscale under extreme conditions.~\cite{Sachdev}  At present the demand for a sub-Kelvin MFM is high for investigating $f$-electron physics as a recent scanning tunneling microscopy study resolves the competition of $f$- and $d$-electrons in heavy fermion systems.~\cite{Aynajian, Allan} In addition, the use of a vector magnet allows further insight into anisotropy in superconductivity such as non-uniform pairing symmetry and magnetic penetration depth.~\cite{xu-95} 

In this Article we report the construction of a $^3$He MFM with a vector field of 2-2-9 T for the $x-y-z$ direction and a base temperature of $T$=300 mK. We calibrate the system's stray field by imaging Abrikosov vortices as a function of field, and demonstrate its low temperature and vector field capabilities by resolving the coexistence of superconductivity and magnetism in a ferromagnetic superconductor at $T$= 500 mK, as well as by imaging the change of a single vortex shape depending on the direction of  the tip magnetization. Our $^3$He MFM system allows investigating phenomena emerging under extreme conditions such as superconductivity in heavy fermions, anisotropic superconductivity, and the interplay of magnetism and superconductivity.

\section{Instrumentation}

\subsection{$^3$He Cryostat insert \& dewar with a vector magnet}

\begin{figure*}
\includegraphics[width=16cm]{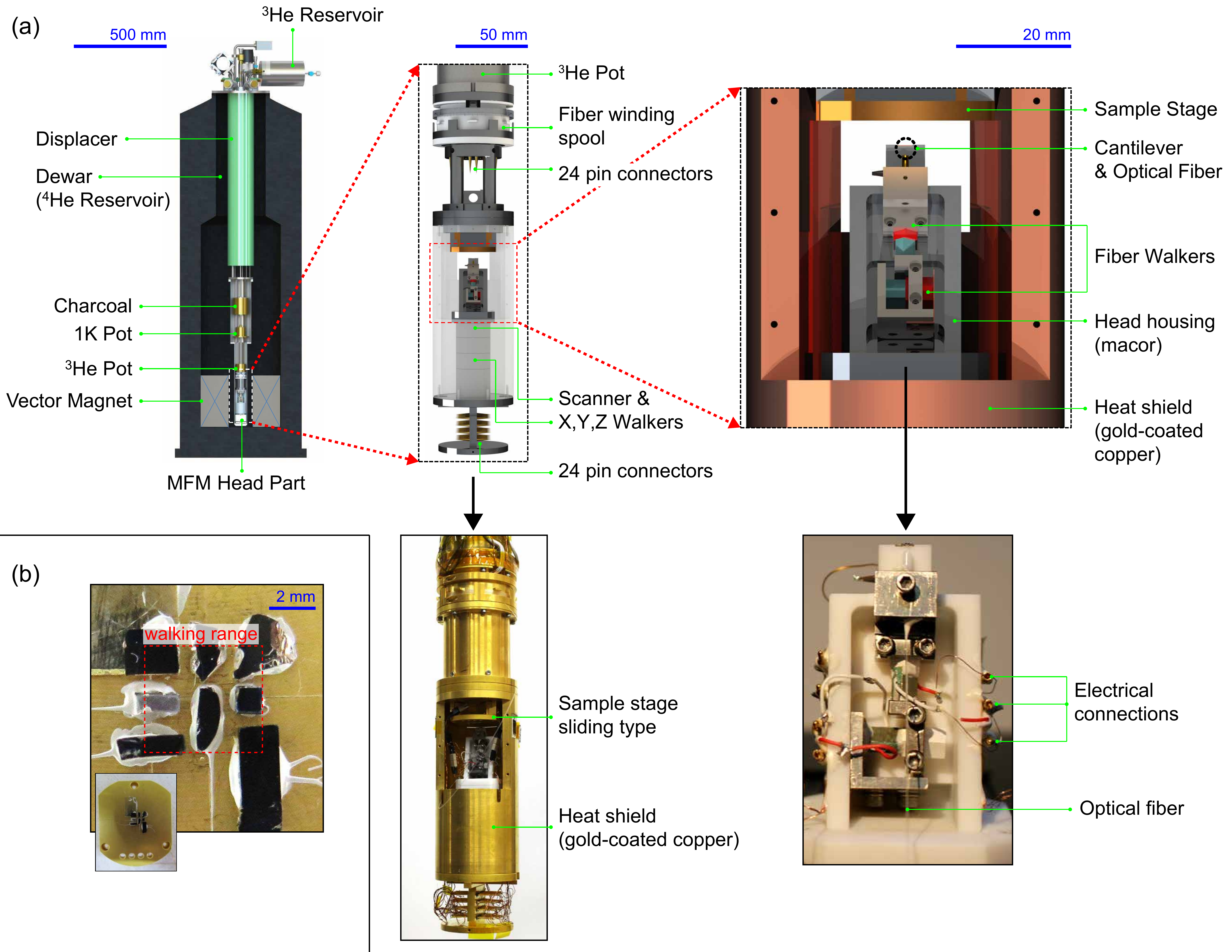}
\caption{(a) Schematic cross-section for the front view of the whole MFM system, a zoom-in on the MFM head, and a zoom-in on the microscope. (b) Top view of the sample stage.\label{sch}}
\end{figure*}

We employ a custom-built $^3$He cryostat and a low-loss dewar with a vector magnet from JANIS [see Fig. \ref{sch}(a)]. The outer jacket of the dewar is evacuated to a pressure lower than $10^{-5}$ Torr prior to cooldown. The liquid $^4$He consumption is about 10 liters a day without magnetic field applied. To minimize the effect of heat radiation, baffles are installed inside the displacer. The heat load from electrical wires has been minimized by winding them around thermal anchors which are attached to the 1-K pot. We control the sample temperature via charcoal, 1-K pot, and $^3$He pot [see Fig. \ref{sch}(a)]. The base temperature ($T$=300 mK) of the $^3$He pot is reached within 4 hours from $T$=4 K and remains for 8 hours if initially all 25 L of $^3$He gas were condensed. The three-axis vector magnet is located at the bottom of the dewar as shown in Fig. \ref{sch}(a) and immersed in liquid $^4$He cryogen. We can apply the vector field of 2-2-9 T in each $x-y-z$ direction with a field homogeneity of $\pm 0.1 \%$ over a sphere of 1 cm diameter at the sample position. The electronics for the vector magnet are supplied by Cryomagnetics Inc., and controlled by a home-made LabVIEW-based program. 

\subsection{MFM head}

The MFM head consists of three parts: (i) the tip stage, (ii) the sample stage, and (iii) the scanner \& $x,y,z$ coarse positioners.
\subsubsection{Tip stage}
The tip stage consists of the cantilever, an optical fiber, and two fiber walkers, as shown in Fig. \ref{sch}(a). It detaches conveniently from the MFM head, allowing an easy replacement of a cantilever. The cantilever is secured via a Be-Cu spring which also establishes an electrical connection for electric scanning probe microscopy modes such as conducting AFM, electric force microscopy, and Kelvin force microscopy. We use an optical interferometric system for measuring the cantilever motion that is directly proportional to the force between the tip and sample. A mechanism to optimize the alignment between the cantilever and the fiber end is of importance for low temperature applications.~\cite{Oral} Therefore, we adopt two compact home-built walkers to control the relative position of the fiber and the tip in the $x$-$z$ direction: $x$ for the cantilever's center in width and $z$ for the fiber-tip distance. The intensity and sensitivity of the MFM signal can be maximized by the following procedure. 
First, the $x$ walker locates the fiber end at the middle of the cantilever to achieve the maximum signal intensity. Then, the $z$ walker regulates the tip-fiber position at the maximum slope of the interference curve to achieve a maximum sensitivity. Considering a sinusoidal waveform for the interference signal, $I(D) = \sin{(2 \pi D/\lambda)}$, the maximum slope of the signal is reached at $\frac{d^2 I(D)}{dD^2} = 0$, where $D$ is the tip-fiber distance and $\lambda$ is the wavelength (1550 nm) of the laser. This condition results in:
\begin{equation}
D = (2n+1)\lambda/8,
\end{equation} 
where $n$ is an integer number. At this point, the sensitivity of the MFM signal ($I$) reaches its maximum, as $\frac{dI}{dD}$ is maximized.~\cite{Rugar} 

\subsubsection{Sample stage}
The sample stage is attached directly to the $^3$He pot via four Cu rods, allowing a good thermal anchoring. A horizontal sliding mechanism for replacing the sample holder ensures a safe and easy mounting procedure. The Au-plated Cu sample holder has a resistive heater element (a 9-$\Omega$ chip resistor for surface mount) and a cernox temperature sensor for an additional temperature control together with the $^3$He-pot control. We can study several samples during a single cooldown with our multi-sample stage [see Fig. \ref{sch}(b)] due to the large walking range of the piezo walkers. This not only avoids lengthy warm-up and cool-down procedures for changing samples, but it also allows to perform comparative studies among the installed samples under controlled conditions, confirming that the geometry of the cantilever, the tip magnetization, etc. remain unchanged between the measurements. The sample holder is grounded electrically to the MFM head [see Fig. \ref{sch}(a)].

\subsubsection{Scanner \& walkers}
In general, a piezo tube scanner, routinely used in scanning tunneling microscope, offers a simple, cost-efficient, and rigid scanning solution in spite of a crosstalk between $z$ and $x-y$ motion as well as a small scan range in the $z$ direction. For magnetic force microscopic applications the large scan area is required to image the $\mu$m-size magnetic domains. We therefore employ a compact, commercially available Attocube scanner (ANSxyz100). Its principle is based on a piezo-driven lever arm mechanism and each axis is decoupled from one another. It offers a scan range of 50$\times$50$\times$24 $\mu$m$^3$ at 300 K and 30$\times$30$\times$15 $\mu$m$^3$ at 4 K, respectively. We use three Attocube walkers (Attocube, ANPx101/RES, ANPz/RES) with a travel distance of 5 mm each for a coarse positioning. An electrical connection to the scanner and walkers is established through a home-built 24-pin connector at the bottom of the head [see Fig. \ref{sch}(a)]. Manganin wires with low thermal and electrical conductivity and Cu wires are used for the scanner and the piezo walkers, respectively, because the high resistance of manganin wires results in a large decay time constant, which is detrimental for the performance of a stick-slip walker. The decay time constant of a driving voltage should be as small as 10 $\mu$sec for a stick-slip walker: For example, the capacitance of the Attocube walker is 5 $\mu$F, requiring the wire resistance is as small as 2 $\Omega$.

\begin{figure}
\includegraphics[width=8cm]{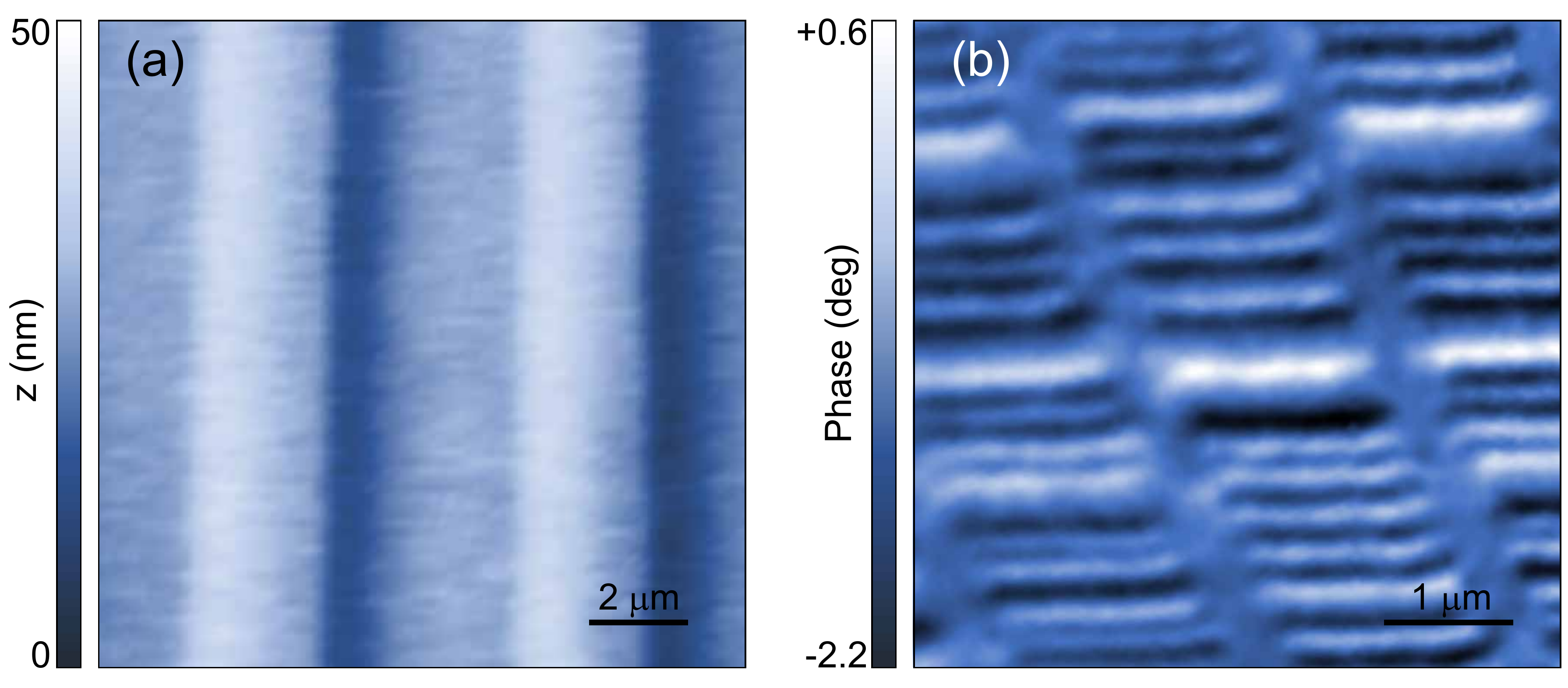}
\caption{(a) Topographic image of the calibration sample from MikroMasch. (b) MFM signal from a disassembled hard disk drive.\label{calibration}}
\end{figure}

We use a standard sample (TGXYZ01/NM from MikroMasch with 20 nm step height and 3 $\mu$m step width) for the scanner calibration. A resulting topographic image obtained at room temperature is shown in Fig. \ref{calibration}(a). In Fig. \ref{calibration}(b) we show the MFM image of a hard disk drive obtained in amplitude-modulation mode at ambient conditions with the tip-lift height of 100 nm. Individual magnetic bits are clearly resolved.

\subsection{Optical interferometric system}

\begin{figure}
\includegraphics[width=8cm]{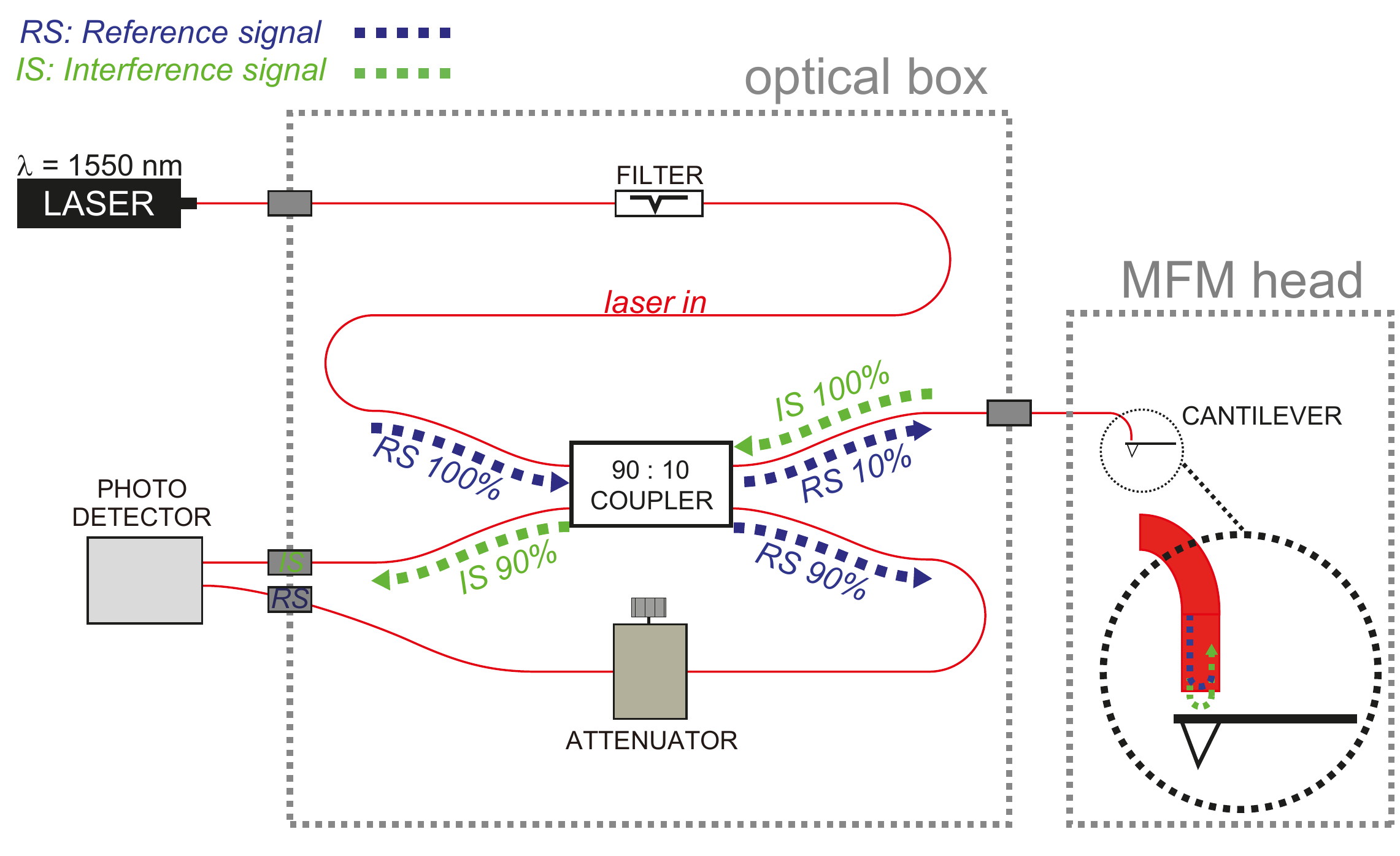}
\caption{Schematics of the optical interferometric system. \label{interferometer}}
\end{figure}

The optical interferometer system provides a stable, low-noise environment for the detection of a cantilever motion.~\cite{Rugar, Smith, Bruland, Oral} We use a 2$\times$2 coupler with an intensity ratio of 9:1 (Gould Fiber Optics). A solid state laser (S3FC1550, THORLABS) of $\lambda$=1550 nm offers a good control of temperature and power. The photon energy of the laser ($\lambda$=1550 nm $\widehat{=}$ 0.8 eV) is smaller than the band gap of a Si cantilever ($E_g$=1.1 eV), and thus transfers less heat than a laser of smaller wavelength. We minimize the laser fluctuation by adjusting the laser power and temperature by monitoring the stability of the interferogram. The adjusted laser power delivered to the cantilever is as small as 40 $\mu$W to avoid a direct cantilever heating. The laser power and temperature on operation are typically 0.34 mW and 23$^\circ$C, respectively. Note that such optimal parameters may vary even with the same laser model. We use angled physical contact (APC) plugs  for all optical connections to minimize undesired reflections from optical connectors though the laser and the photo detector have built-in physical contact (PC) plugs. Therefore, we use a custom spec adaptor to connect PC with APC. The schematic diagram for the interferometer system is depicted in Fig. \ref{interferometer}. The incident laser beam passes through an optical isolator (Thorlabs IO-H-1550APC) to discriminate any side bands, and is split into two beams via an optical coupler. 90\% of the incident beam goes into  the photo detector (new focus model 2117) through an attenuator and serves as a reference signal (blue dotted lines in Fig. \ref{interferometer}). The remaining 10\% are fed into the microscope head, where they partly transmit through the cleaved fiber end, and reflect at the cantilever. The remaining part of the beam is reflected at the fiber end, and interferes with the reflected light from the cantilever backside within the fiber; this interference signal is fed into the coupler (see Fig. \ref{interferometer}). Both the reference and the interference signal are fed into the balanced photo detector, which allows the DC part of the interference signal to be zero since it is subtracted from the interference. Thus, the removal of the DC signal results in a wide dynamic span of an amplifier allowing an increase of the DC gain as well as the subtraction of DC laser fluctuations. The interference intensity measured by a photo detector is as follows:
\begin{equation}
I = I_1 - I_2 \cos{\left(\frac{4\pi D}{\lambda}\right)},
\end{equation}
where $I_1$ is the laser power and $I_2$ is proportional to $I_1$ multiplied by the optical visibility of the system, $D$ is the fiber-cantilever distance, and  $\lambda$ is the wavelength of the laser.~\cite{Nazaretski} For a fixed wavelength, the intensity change due to the cantilever motion is the most sensitive when $4\pi D/\lambda$ is close to a multiple of $\pi/2$, because the variation in intensity due to the change of $D$ is maximum at these points: For example, $\frac{dI}{dD}$ is maximized when $4\pi D/\lambda$ is a multiple of $\pi/2$. Therefore we regulate the fiber-cantilever distance at these crossing points by the fiber piezo.

\section{Characterization and performance}

\subsection{Vibration isolation and noise characterization}

\begin{figure}
\includegraphics[width=8cm]{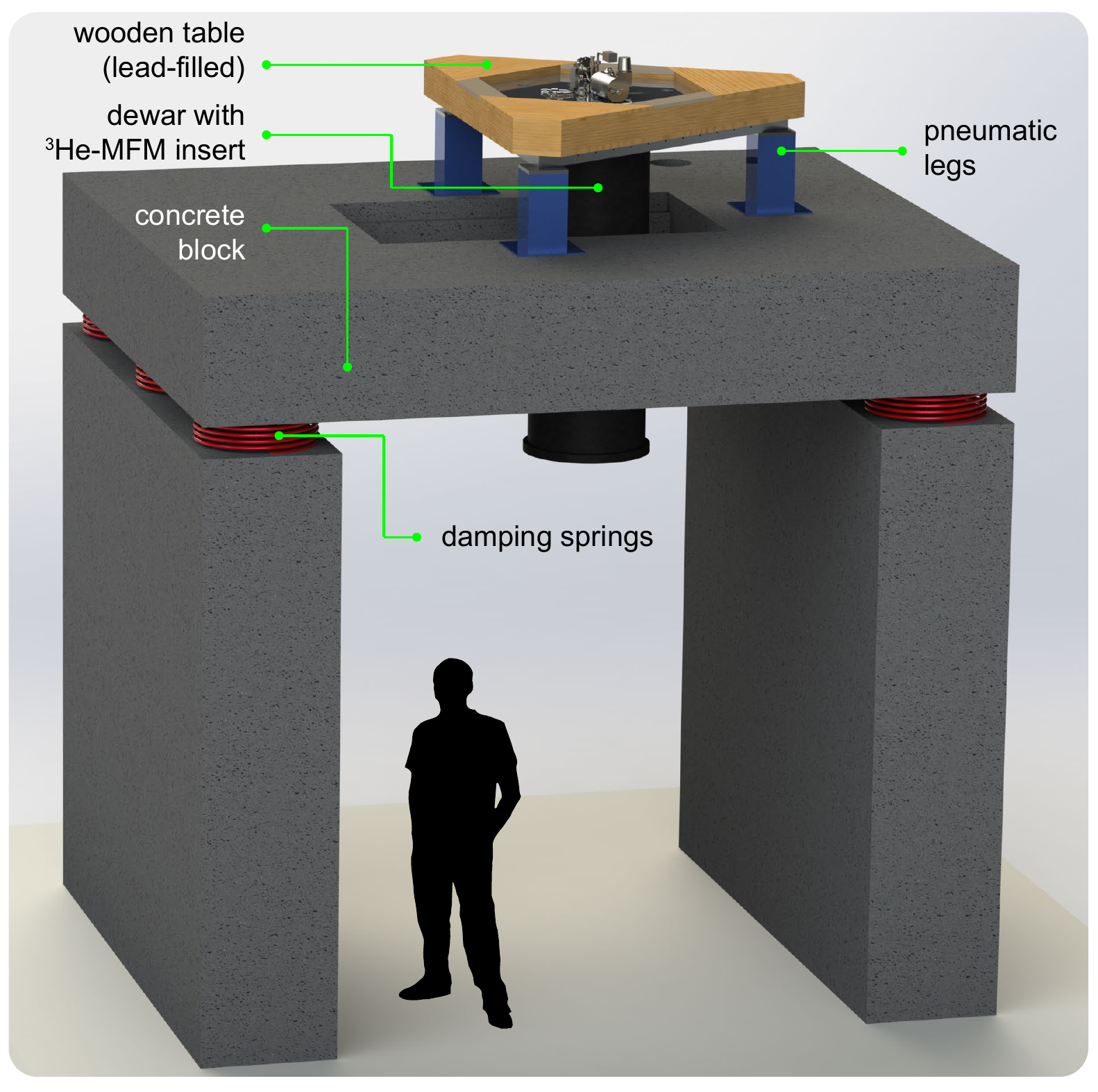}
\caption{Schematics of the vibration isolation laboratory.
\label{vibration}}
\end{figure}

In order to minimize the mechanical noise, our microscope will be installed into a newly constructed vibration isolation laboratory at the Institute for Basic Science in Pohang, as shown in Fig. \ref{vibration}. A 20-ton concrete block resting on the heavy duty damping springs serves as the base for our system. Three pneumatic legs standing on the concrete block carry a home-built triangular wooden table filled with lead bricks of 800 kg to further lower the system's resonance frequency. The $^3$He cryostat insert is fixed on the table, and the dewar with a vector magnet moves up and down for experiments through the rectangular pit.

We characterize the thermal noise level of our microscope as follows:~\cite{Albrecht}
\begin{equation}
\delta f^{min} = \sqrt{\frac{f_0k_BTB}{2\pi kQA^2}},
\end{equation}
where $f_0$, $k_B$, $T$, $B$, $k$, $Q$, and $A$ are resonance frequency, Boltzmann constant, temperature, measurement bandwidth, spring constant, quality factor, and oscillation amplitude, respectively. Based on our typical operation parameters ($f_0=75195$ Hz, $B=100$ Hz, $k=2.8$ N/m, $Q=48554$, and $A=4.8$ nm) at $T$=4.2 K, the minimum detectable frequency shift is $\delta f^{min}=4.6$ mHz, and the corresponding minimal detectable force gradient is:
\begin{equation}
\left\vert \frac{\partial F_z^{min}}{\partial z} \right\vert = 2k \cdot \frac{\delta f^{min}}{f_0} = 3.4 \cdot 10^{-7} \mathrm{N/m}.
\end{equation}
The measured background, as shown by the red profile line in Fig. \ref{nb}(a), varies by about $7.4 \cdot 10^{-7}$ N/m, which only deviates by a factor of 2 from the thermodynamic noise limit. Additional noise fluctuations can be attributed to electrical noise, mechanical vibrations, and possible magnetic inhomogeneities at the measurement position. The MFM signal magnitude of an Abrikosov vortex in the Nb film [black profile line in Fig. \ref{nb}(a)] exceeds the noise level by a factor of about 40, which is comparable to the previous report.~\cite{Nazaretski}

\subsection{MFM results}
Here we present two experimental results demonstrating the performance of our home-built $^3$He MFM. The images of Abrikosov vortices obtained in a Nb film highlight the vector magnet capability. The simultaneous image of ferromagnetism and superconductivity at $T$=500 mK in ErNi$_2$B$_2$C proves the sub-Kelvin operation.

\subsubsection{Vector magnet experiments}

\begin{figure}
\includegraphics[width=8cm]{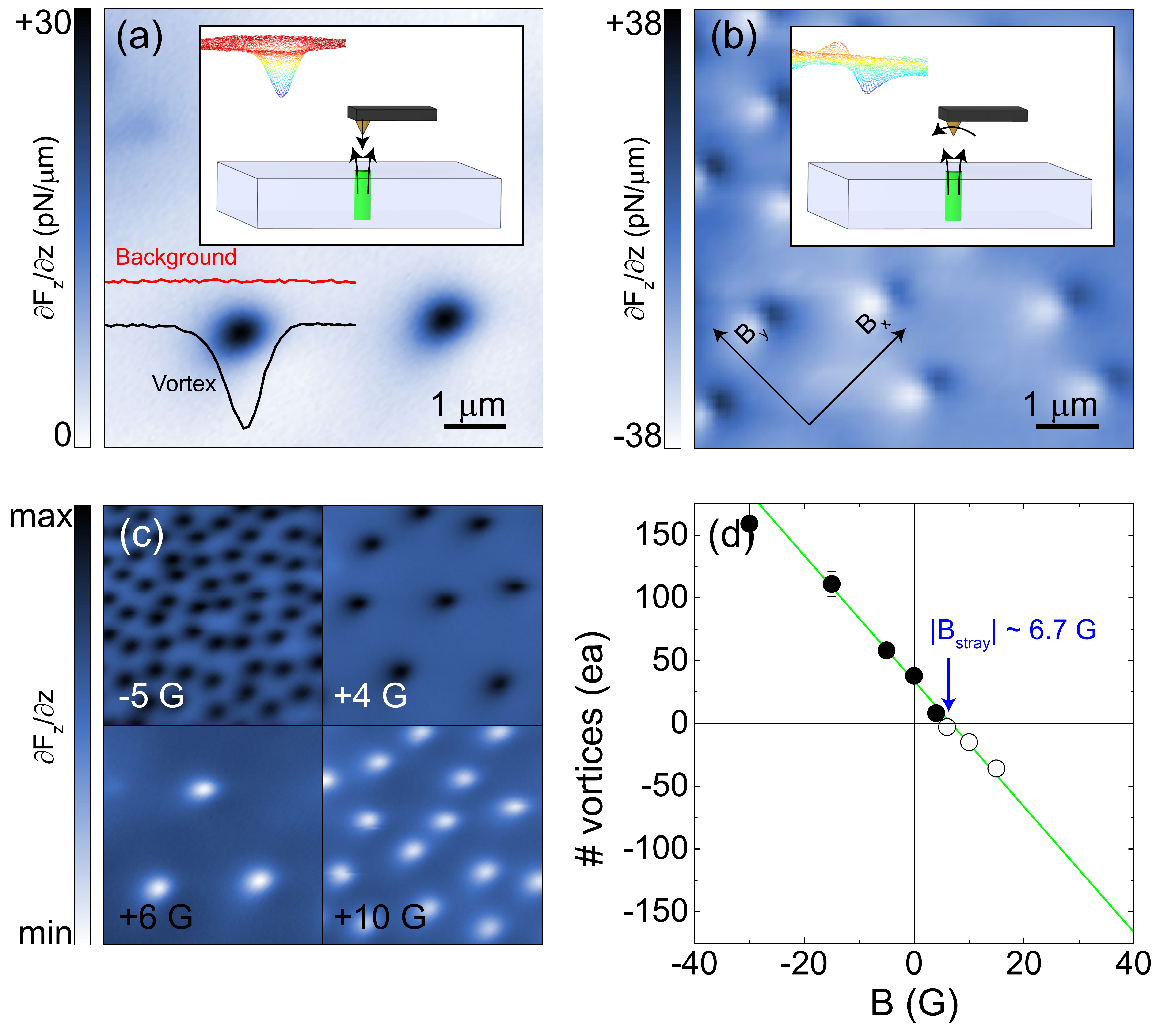}
\caption{(a) Abrikosov vortices in a superconducting Nb film imaged with an out-of-plane magnetized MFM tip. Line profiles of the vortex and the background are presented as black and red solid lines. (b) Vortices imaged at the same sample position with an in-plane magnetized MFM tip. The insets in (a) and (b) indicate the magnetic field direction of the tip and vortex. (c) Four representative MFM images obtained at different magnetic fields. (d) Number of vortices on a $10 \times 10$ $\mu$m$^2$ area as a function of the out-of-plane applied magnetic field. All images were taken at $T=4.2$ K and with a tip-sample distance of 300 nm.\label{nb}}
\end{figure}

The images of vortices in Fig. \ref{nb}, obtained from a Nb film, show the vector magnet capability of our MFM. The Nb thin film was grown on a Si substrate via electron beam deposition with a film thickness and critical temperature of 300 $\pm$ 5 nm and $T_c=8.8$ K, respectively. The tip was magnetized along the $z$ direction, perpendicular to the surface of the Nb film, by applying a magnetic field of 10 kOe larger than the tip coercive field ($\sim$ 300 Oe). We field-cool the sample from $T=12$ K (above $T_c$) in the presence of several Oersted. Due to local pinning centers the penetrating flux quanta can be trapped within the scan frame, while the sample remains in the pure Meissner state. The resulting MFM image is shown in Fig. \ref{nb}(a): the inset sketches the field lines from a single vortex. The vortex image with an in-plane tip magnetization is shown in Fig. \ref{nb}(b) after polarizing the tip magnetization along the in-plane direction by a vector magnet. The scan areas in Fig. \ref{nb}(a) and (b) are identical. The MFM signal of a vortex resembles a dipole, which is in contrast to the monopole signal of an out-of-plane tip magnetization directions [see Fig. \ref{nb}(a)]. The two different vortex images result from the two different tip magnetizations. In general, the force gradient is positive for an attractive interaction between the tip and vortex as it is negative for a repulsive interaction. The tip magnetized with an in-plane field experiences both positive and negative force gradients as it crosses a single vortex, which results in dipole-like vortex as shown in Fig. \ref{nb}(b). This result displays a good demonstration of a vector field effect.
 
Imaging vortices as a function of applied field allows us to estimate the stray field of our setup. The vortex images at four different magnetic fields are shown in Fig. \ref{nb}(c). Note that the polarity of the vortices changes the sign with field, as indicated by the MFM contrast. The number of vortices per scan frame as a function of fields is shown in Fig. \ref{nb}(d). The experimental flux quantum $\Phi_{exp}$ can be obtained from Fig. \ref{nb}(d) as follows: $n=\frac{1}{\Phi_{exp}A}B+B_{stray}$, where $n$ is a number of vortices for a scan area of $A$, $B_{stray}$ is the stray field. We find a stray field of $B_{stray}$= +6.7 G, and for a given area of $A$=100 $\mu$m$^2$, we obtain $\Phi_{exp} = 20.02 \pm 0.84$ G$\mu$m$^2$, which is in good agreement with the theoretical quantum flux of $\Phi_{0}=h/2e\approx$20.67 G$\mu$m$^2$.

\subsubsection{Ultra-low temperature experiments}
\begin{figure}
\includegraphics[width=8cm]{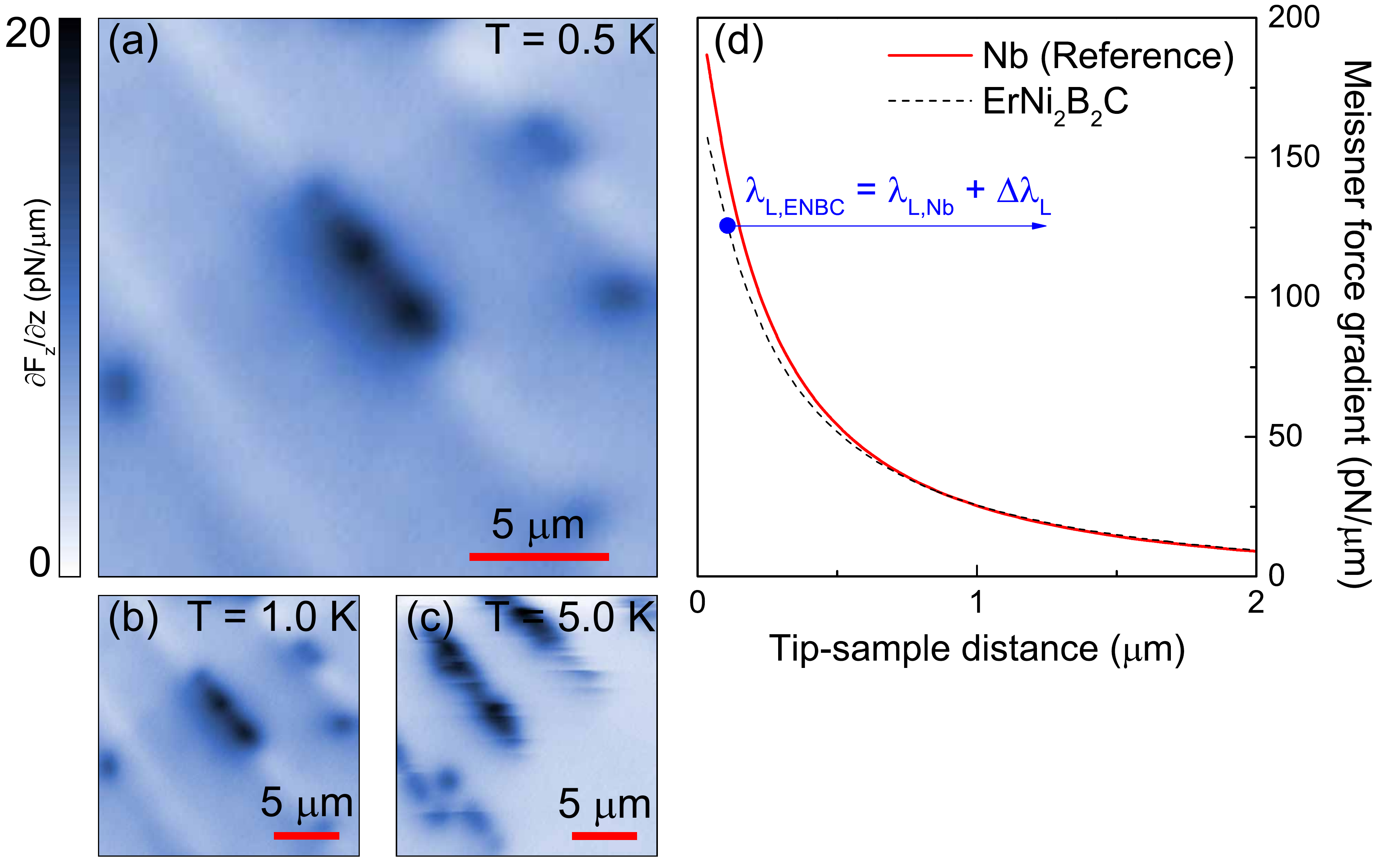}
\caption{(a) MFM image of the magnetic superconductor ErNi$_2$B$_2$C, obtained at $T=500$ mK. (b) and (c): Subsequent images obtained on the same scan area at 1.0 K and 5.0 K, respectively. All images were taken with a tip-sample distance of 700 nm. (d) Comparison of Meissner force curves from ErNi$_2$B$_2$C at 500 mK (dashed black curve) and a reference sample (Nb, solid red curve) to determine $\lambda_L$.~\cite{Wulferding} \label{ENBC}}
\end{figure}

In order to demonstrate the low temperature capability, we investigate magnetism in ErNi$_2$B$_2$C, a magnetic superconductor,  at the sub-Kelvin temperature.~\cite{Wulferding} It turns superconducting below $T_c = 10.5$ K and develops a weak ferromagnetic order below $T_{WFM} = 2.3$ K within the superconducting phase. The MFM image shown in Fig. \ref{ENBC}(a) was obtained at $T=500$ mK, and clearly shows magnetic stripes as well as Abrikosov vortices (dark spots), indicating a coexistence of magnetism and superconductivity. At temperatures higher than $T_{WFM}$ the bright stripes vanish as shown in Figs. \ref{ENBC}(b) and (c), signaling the fading of the magnetic order upon reaching the ferromagnetic transition. At elevated temperatures the vortices tend to be mobile by overcoming local pinning potentials, which allows a manipulation of vortices via the tip magnetic moment upon scanning. This is evident from the smeared-out vortex lines in Fig. \ref{ENBC}(c).
The absolute values of the local London penetration depth ($\lambda_{\mathrm{L}}$) can be obtained through a comparative Meissner technique, as shown in Fig. \ref{ENBC}(d).~\cite{jeehoon1}  When the sample is in the superconducting state, the magnetic tip experiences a  Meissner force which grows algebraically with decreasing the tip-sample distance. Using a comparative method, one can extract the absolute value of $\lambda_{\mathrm{L}}$ by overlaying the Meissner force curve of the sample (dashed black curve) with that of a Nb reference sample (solid red curve), $\lambda_{\mathrm{L,Nb}} = 110$ nm. The addition of the shifted distance of 100 nm to $\lambda_{\mathrm{L,Nb}}$ leads to a penetration depth in ErNi$_2$B$_2$C of $\lambda_{\mathrm{L,ENBC}} = 210$ nm. This straightforward method to measure $\lambda_{\mathrm{L}}$, one of the two intrinsic superconducting parameters, at sub-Kelvin temperatures under a vector field allows further investigation of exotic superconductivity together with anisotropy. 

\section{Summary}

In summary, we constructed a $^3$He MFM operating at a base temperature of 300 mK under vector magnetic fields of {\bf H}$_{xyz}$=2-2-9 T. We employed a low-noise fiber interferometer system with a noise level close to its thermodynamic limit. The sub-Kelvin and vector field capabilities of the apparatus were demonstrated by imaging the shape change of the Abrikosov vortices from monopole to dipole according to the rotation of the tip magnetization from out-of-plane to in-plane in a Nb film and by resolving a coexistence of magnetism and superconductivity in ErNi$_2$B$_2$C at $T$=500 mK. Our $^3$He  MFM opens the door to investigate magnetism in exotic superconducting phenomena emerging at sub-Kelvin temperatures, such as magnetically mediated superconductivity in heavy fermion systems and the nonlinear Meissner effect with an anisotropic pairing symmetry. It furthermore allows us to study the anisotropy of unconventional magnetic domains in a vector magnetic field.

\begin{acknowledgments}
This work was supported by the Institute for Basic Science (IBS), Grant No. IBS-R014-D1.
\end{acknowledgments}


\begin{thebibliography}{9}

\bibitem{Martin}
Y. Martin and H. K. Wickramasinghe, Appl. Phys. Lett. \textbf{50}, 1455 (1987).

\bibitem{LSMO}
J. Jeong, I. Yang, J. Yang, O. E. Ayala-Valenzuela, D. Wulferding, J.-S. Zhou, J. B. Goodenough, A. de Lozanne, J. F. Mitchell, N. Leon, R. Movshovich, Y. H. Jeong, H. W. Yeom, and J. Kim, Phys. Rev. B \textbf{92}, 054426 (2015).

\bibitem{Liebmann}
M. Liebmann, A. Schwarz, S. M. Langkat, and R. Wiesendanger, Rev. Sci. Instrum. \textbf{73}, 3508 (2002).

\bibitem{Schwarz2}
A. Schwarz, M. Liebmann, U. Kaiser, R. Wiesendanger, T. W. Noh, and D. W. Kim, Phys. Rev. Lett. \textbf{92}, 077206 (2004).

\bibitem{Moser}
A. Moser, H. J. Hug, I. Parashikov, B. Stiefel, O. Fritz, H. Thomas, A. Baratoff, H.-J. G\"{u}ntherodt, and P. Chaudhari, Phys. Rev. Lett. \textbf{74}, 1847 (1995).

\bibitem{Volodin}
A. Volodin, K. Temst, C. Van Haesendonck, Y. Bruynseraede, M. I. Montero, and I. K. Schuller, Europhys. Lett. \textbf{58}, 582 (2002).

\bibitem{Schwarz}
A. Schwarz, U. H. Pi, M. Liebmann, R. Wiesendanger, Z. G. Khim, and D. H. Kim, Appl. Phys. Lett. \textbf{88}, 012507 (2006).

\bibitem{Roseman2}
M. Roseman and P. Gr\"{u}tter, J. Appl. Phys. \textbf{91}, 8840 (2002).

\bibitem{Pi}
U. H. Pi, A. Schwarz, M. Liebmann, R. Wiesendanger, Z. G. Khim, and D. H. Kim, Phys. Rev. B \textbf{73}, 144505 (2006).

\bibitem{Luan}
L. Luan, O. M. Auslaender, D. A. Bonn, R. Liang, W. N. Hardy, and K. A. Moler, Phys. Rev. B \textbf{79}, 214530 (2009).

\bibitem{EWJ}
E. W. J. Straver, J. E. Hoffman, O. M. Auslaender, D. Rugar, and K. A. Moler, Appl. Phys. Lett. \textbf{93}, 172514 (2008).

\bibitem{Nb}
I. Yang, J. Yang, D. Wulferding, S. Chung, H. Yeom, K. Kim, and J. Kim (unpublished).

\bibitem{Milde}
P. Milde, D. K\"{o}hler, J. Seidel, L. M. Eng, A. Bauer, A. Chacon, J. Kindervater, S. M\"{u}hlbauer, C. Pfleiderer, S. Buhrandt, C. Sch\"{u}tte, and A. Rosch, Science \textbf{340}, 1076 (2013).

\bibitem{Hug}
H. J. Hug, Th. Jung, and H.-J. G\"{u}ntherodt, Rev. Sci. Instrum. \textbf{63}, 3900 (1992).

\bibitem{Euler}
R. Euler, U. Memmert, and U. Hartmann, Rev. Sci. Instrum. \textbf{68}, 1776 (1997).

\bibitem{Roseman}
M. Roseman and P. Gr\"{u}tter, Rev. Sci. Instrum. \textbf{71}, 3782 (2000).

\bibitem{Chuang}
T.-M. Chuang and A. de Lozanne, Rev. Sci. Instrum. \textbf{78}, 053710 (2007).

\bibitem{Nazaretski}
E. Nazaretski, K. S. Graham, J. D. Thompson, J. A. Wright, D. V. Pelekhov, P. C. Hammel, and R. Movshovich, Rev. Sci. Instrum. \textbf{80}, 083704 (2009).

\bibitem{Pelekhov}
D. V. Pelekhov, J. B. Becker, and G. Nunes, Rev. Sci. Instrum. \textbf{70}, 114 (1999).

\bibitem{Ozgur}
\"{O}. Karci, M. Dede, and A. Oral, Rev. Sci. Instrum. \textbf{85}, 103705 (2014).

\bibitem{Thalmeier}
P. Thalmeier, G. Zwicknagl, G. Sparn and F. Steglich, ``Superconductivity in Heavy Fermion Compounds'', in ``Frontiers in Superconducting Materials'', ed. A. V. Narlikar, (Springer Verlag 2005), p. 109.

\bibitem{Sachdev}
S. Sachdev, Nature Physics \textbf{4}, 173 (2008).

\bibitem{Aynajian}
P. Aynajian, E. H. da Silva Neto, A. Gyenis, R. E. Baumbach, J. D. Thompson, Z. Fisk, E. D. Bauer, and A. Yazdani, Nature \textbf{486}, 201 (2012).

\bibitem{Allan}
M. P. Allan, F. Massee, D. K. Morr, J. Van Dyke, A. W. Rost, A. P. Mackenzie, C. Petrovic, and J. C. Davis, Nature Physics \textbf{9}, 468 (2013).

\bibitem{xu-95}
D. Xu, S. K. Yip, and J. A. Sauls, Phys. Rev. B \textbf{51}, 16233 (1995).

\bibitem{Oral}
A. Oral, R. A. Grimble, H. \"{O}. \"{O}zer, and J. B. Pethica, Rev. Sci. Instrum. \textbf{74}, 3656 (2003).

\bibitem{Rugar}
D. Rugar, H. J. Mamin, R. Erlandsson, J. E. Stern, and B. D. Terris, Rev. Sci. Instrum. \textbf{59}, 2337 (1988).

\bibitem{Smith}
D. D. Smith, J. A. Marohn, and L. E. Harrell, Rev. Sci. Instrum. \textbf{72}, 2080 (2001).

\bibitem{Bruland}
K. J. Bruland, J. L. Garbini, W. M. Dougherty, S. H. Chao, S. E. Jensen, and J. A. Sidles, Rev. Sci. Instrum. \textbf{70}, 3542 (1999).

\bibitem{Albrecht}
T. R. Albrecht, P. Gr\"{u}tter, D. Horne, and D. Rugar, J. Appl. Phys. \textbf{69}, 668 (1991).

\bibitem{Wulferding}
D. Wulferding, I. Yang, J. Yang, M. Lee, H. C. Choi, S. L. Bud'ko, P. C. Canfield, H. W. Yeom, and J. Kim, Phys. Rev. B \textbf{92}, 014517 (2015).

\bibitem{jeehoon1}
J. Kim, L. Civale, E. Nazaretski, N. Haberkorn, F. Ronning, A. S. Sefat, T. Tajima, B. H. Moeckly, J. D. Thompson, and R. Movshovich, Supercond. Sci. Technol. \textbf{25}, 112001 (2012).

\end{thebibliography}
\end{document}